\documentclass[sigconf, screen]{acmart}

\usepackage{newfloat}
\usepackage{amsthm}
\usepackage{array}
\usepackage{graphicx}
\usepackage{xcolor}
\usepackage{subfigure}
\usepackage{colortbl} 
\usepackage{microtype}
\usepackage{algpseudocode}
\usepackage{amsfonts}
\usepackage{threeparttable}
\usepackage{pifont} 
\usepackage{enumitem}
\usepackage{bm} 
\usepackage{amsmath}
\usepackage{booktabs,multirow,multicol}
\usepackage{makecell} 

\newcommand{\ul}[1]{\underline{#1}}
\newtheorem{definition}{Definition}

\providecommand{\ul}[1]{\underline{#1}} 

\AtBeginDocument{%
  }

\setcopyright{acmlicensed}
\copyrightyear{2026}
\acmYear{2026}
\acmDOI{}
\acmConference[MM'26]{In Proceedings of the 34rd ACMInternational Conference on Multimedia}{November 10--14,
  2026}{Rio de Janeiro, Brazil}
\acmISBN{978-1-4503/2026/06}

\begin{document}

\title{EviRank: Evidence-Based Confidence Estimation for LLM-Based Ranking}
\author{Meng Yan}
\affiliation{%
  \institution{Xidian University}
  \city{Xi'an}
  \country{China}}
\orcid{0000-0001-8478-4823}
\email{mengyan@stu.xidian.edu.cn}

\author{Cai Xu}
\affiliation{%
  \institution{Xidian University}
  \city{Xi'an}
  \country{China}}
\orcid{0000-0002-7191-7348}
\email{cxu@xidian.edu.cn}

\author{Xujing Wang}
\affiliation{%
 \institution{Xidian University}
 \city{Xi'an}
 \country{China}}
\orcid{0009-0003-5336-8845}
\email{xjwong@stu.xidian.edu.cn}

\author{Ziyu Guan}
\authornote{Corresponding author}
\affiliation{%
  \institution{Xidian University}
  \city{Xi'an}
  \country{China}}
\orcid{0000-0003-2413-4698}
\email{zyguan@xidian.edu.cn}

\author{Wei Zhao}
\affiliation{%
  \institution{Xidian University}
  \city{Xi'an}
  \country{China}}
\orcid{0000-0002-9767-1323}
\email{ywzhao@mail.xidian.edu.cn}

\renewcommand{\shortauthors}{Trovato et al.}

\begin{abstract}
Large Language Models show promise for recommendation,  but they raise reliability concerns due to limited domain coverage and inherent stochasticity.
Existing uncertainty quantification methods persist two fundamental challenges: 
(1) the global confidence score designed for question answering fails to reveal which positions are unreliable in ranking list; 
(2) fine-grained confidence extracted from model internals exhibits uniformly low values across all positions, making it impossible to filter unreliable predictions.

To tackle the challenges, we propose an evidence-based confidence estimation for LLM-based ranking (EviRank). 
We extract three complementary evidences from a single forward pass and aggregate them via reliable opinion aggregation. 
Furthermore,  we recognize that ranking positions are inherently unequal, and introduce a position-aware calibration. Lastly, the calibrated confidence guides ranking optimization. Experiments on three datasets demonstrate that our method achieves state-of-the-art performance on both recommendation and uncertainty quantification.
Our source code and prompt template is available at \href{https://anonymous.4open.science/r/EviRank-CDE0}{
 https://anonymous.4open. science/r/EviRank-CDE0}.
\end{abstract}





\maketitle
\vspace{-0.2mm}
\section{Introduction}

Large Language Models (LLMs) are emerging as a transformative paradigm in recommender systems. Leveraging powerful natural language understanding capabilities, LLMs can unify user preferences, item descriptions, and contextual signals within a textual semantic space, directly generating personalized ranked lists through carefully designed prompts along with natural language explanations~\cite{survey}. Recent work such as ChatRec~\cite{chatrec}, A-LLMRec~\cite{allmrec}, and TransRec~\cite{transrec} has demonstrated promising potential for zero-shot and explainable recommendation.

\begin{figure}[t]
\centering
\includegraphics[width=0.48\textwidth]{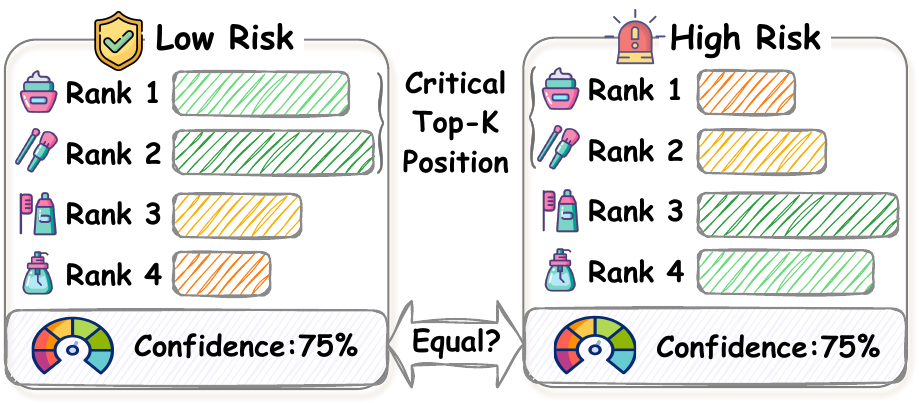} 
\caption{Illustration of the limitations of global confidence in ranking. Both lists share a same global confidence score, yet they represent drastically different risk profiles.}
\label{fig:intro}
\end{figure}

However, LLMs are predominantly trained on general-purpose text corpora with limited coverage of domain-specific items (e.g., newly released movies, specialized books)~\cite{challenge1}. When candidate sets contain items unseen during training, the model lacks sufficient knowledge to reliably assess their alignment with user preferences. Furthermore, influenced by decoding strategies and probability distributions, LLMs exhibit inherent stochasticity~\cite{challenge2}. Even with identical user histories and candidate sets, multiple runs may yield different rankings. This raises a critical question: 
\textbf{\textit{how confident can we be that the ranked lists generated by LLMs are reliable?}}

A straightforward idea is to directly apply existing LLM uncertainty quantification methods, these approaches can be categorized into two paradigms. The first is output-based methods: self-verbalized confidence~\cite{verb.1S} prompts the model to articulate an overall confidence score, while semantic entropy~\cite{semantic} measures uncertainty through the consistency of multiple generated outputs. The second paradigm involves internal-based methods, such as logit lens~\cite{label}, probing classifiers~\cite{probe}, and sparse auto-encoders~\cite{sparse}, which decode uncertainty signals from intermediate layer representations. Both paradigms prove effective in question answering or short-form text generation. For example, a question like "\textit{What is the capital of France?}", responding "\textit{Paris, confidence: 95\%}" adequately conveys the model's certainty. 
However, for ranking tasks in recommendation, a single global uncertainty score is far from sufficient. As illustrated in Figure~\ref{fig:intro}, two recommendation results both display 75\% overall confidence, yet in the left result uncertainty concentrates at the tail of the list, while in the right result uncertainty is heavily concentrated at the critical top positions. Consequently, under identical uncertainty scores, the latter poses substantially higher actual risk.

\begin{figure}[t]
\centering
\includegraphics[width=0.47\textwidth]{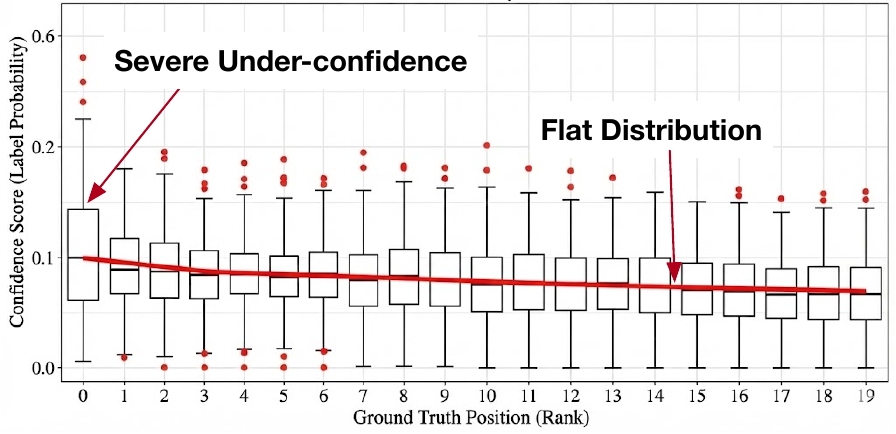} 
\caption{Confidence distribution (label probability) across ground truth positions on Amazon Grocery.}
\label{fig:intro2}
\end{figure}

Furthermore, we explore more fine-grained methods based on internal information for measuring uncertainty. Ideally, these methods should reflect confidence strength at different positions.
We employ Label Probability~\cite{label} to measure position-level confidence on the Amazon Grocery dataset~\cite{amazon} and analyze the confidence distribution corresponding to ground-truth items at different ranking positions (Figure~\ref{fig:intro2}). The results reveal a calibration deficiency. Confidence scores are uniformly low across all positions. Even for ground-truth items ranked at position 0, which ostensibly represents the model's most confident recommendations, the confidence predominantly falls within the [0.05,0.15], indicating severe under-confidence. Moreover, the confidence differential across positions is negligible. This near-flat distribution renders it practically impossible to filter low-quality recommendations through threshold-based mechanisms.


Based on the above considerations, we propose an \textbf{evi}dence-based confidence estimation for LLM-based \textbf{rank}ing (\textbf{\textsc{EviRank}}).
Specifically, we extract three complementary evidences from a single forward pass of the LLM, then aggregate them through a reliable opinion aggregation to obtain robust belief masses. 
Since ranking positions in a ranked list are inherently unequal, we introduce a position-aware calibration by incorporating position importance.
Finally, we leverage a confidence-weighted scoring function to guide reranking, thereby providing more reliable recommendations.

 Overall, our contributions are as follows:
\begin{itemize}
\item We extract semantic, attention, and output evidences from a single forward pass and estimate position-level confidence via reliable opinion aggregation.
 
\item We identify the confidence calibration deficiency, and then introduce a position-aware calibration that incorporates the inherent importance of ranking positions.

\item We conduct extensive experiments on three public real-world datasets, verifying that \textsc{EviRank} achieves state-of-the-art performance in both recommendation and uncertainty quantification.
\end{itemize}

\section{Related Work}
\subsection{LLM-based Recommendation}

Large language models have gained significant attention in recommender systems due to their powerful natural language understanding and generation capabilities~\cite{survey}. Early efforts adapt language model architectures for recommendation tasks, demonstrating improvements over conventional collaborative filtering methods~\cite{tallrec,p5}. With the emergence of powerful pre-trained LLMs~\cite{llama,qwen2.5}, subsequent research explores their effectiveness for zero-shot~\cite{zero} and few-shot~\cite{few1,few2} recommendation without task-specific training. Recent work further fine-tunes LLMs with instruction tuning on recommendation datasets to bridge the gap between general language understanding and recommendation~\cite{transrec,chatrec}.
However, LLMs trained on general-purpose corpora have limited coverage of domain-specific items, and their inherent stochasticity can lead to inconsistent rankings~\cite{challenge1,challenge2}. These reliability concerns motivate the need for confidence estimation in LLM-based recommendation.

\begin{figure*}[t]
\centering
\includegraphics[width=1\textwidth]{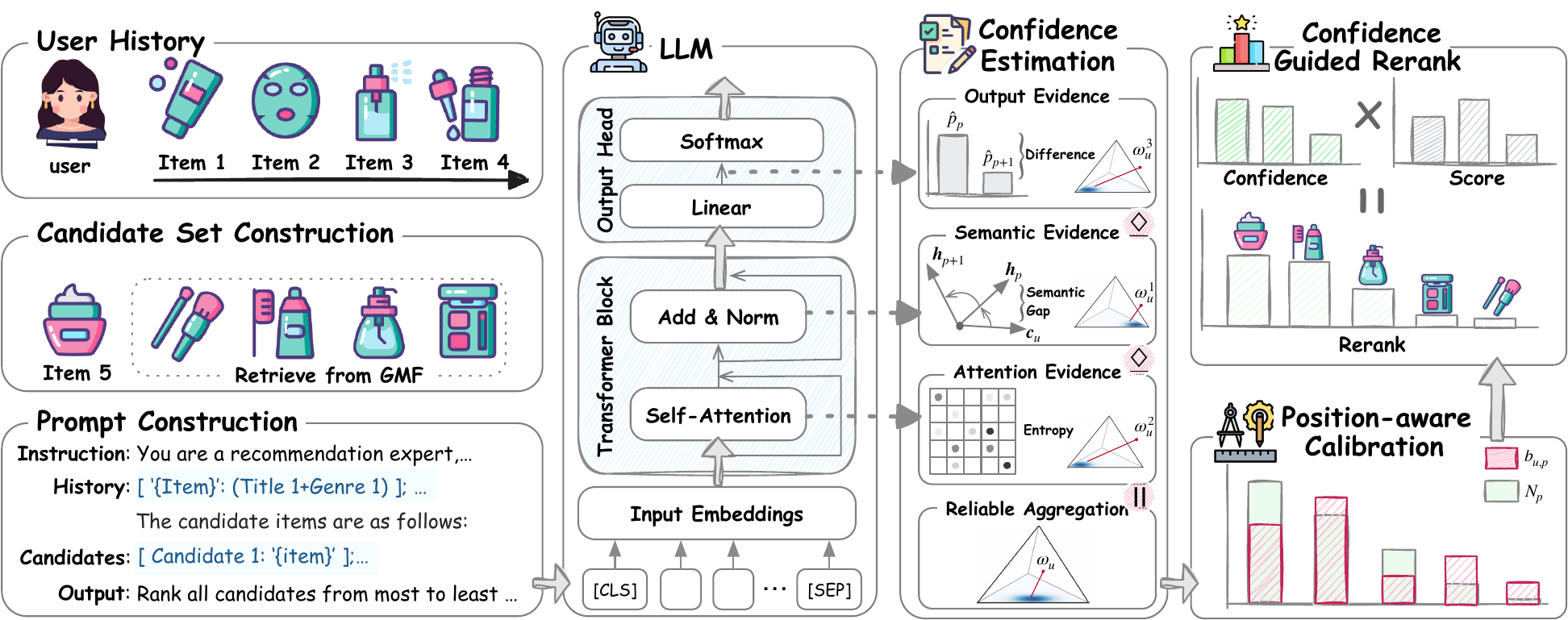} 
\caption{Illustration of \textsc{EviRank}. The model comprises three key components: (1)~evidence-based confidence estimation; (2)~position-aware calibration; (3)~confidence-guided reranking. }
\label{fig:model}
\end{figure*}
\vspace{-0.2cm}
\subsection {\textbf{Uncertainty in LLM}}
Uncertainty quantification for LLMs has been extensively studied, primarily following two methodological paradigms: \textbf{(1) output-based methods}. LLM-generated responses elicit the model to articulate its own confidence scores~\cite{verb.1S}. Sampling-based methods~\cite{sample_uncertainty, semantic} measure consistency across multiple generated outputs to quantify uncertainty; 
\textbf{(2) internal-based methods}. These approaches decode uncertainty signals from intermediate layer representations rather than final outputs. Representative methods include logit lens~\cite{label}, probing classifiers~\cite{probe}, and sparse auto-encoders~\cite{sparse}, which extract uncertainty information from hidden states at various layers.
Both paradigms prove effective for conversation and question answering. However, applying these methods to LLM-based recommendation faces fundamental challenges. First, a global confidence score fails to capture position-level reliability in ranking. Second, we identify calibration bias from fine-grained methods. We thereby propose position-aware confidence estimation for LLM-based recommendation.
\vspace{-0.2cm}
\section{Method}
In this section, we present our method \textsc{EviRank}. We first formalize the problem setting (\ref{Sec:pre}), then introduce our evidence-based confidence estimation (\ref{Sec:conf}), followed by the position-aware calibration (\ref{Sec:cali}), and finally describe the confidence-guided reranking (\ref{Sec:opti}). Figure~\ref{fig:model} illustrates the overall architecture.
\vspace{-0.2cm}
\subsection{Problem Formulation}
\label{Sec:pre}
Let $\mathcal U$ and $\mathcal I$ represent the sets of users and items, respectively. For each user $u \in \mathcal U$, we represent the interaction history as a chronological item sequence $\mathcal H_u = [i_1,i_2, \ldots ,i_{|\mathcal H_u|}]$. Each item $i \in \mathcal I$ is described by a textual representation $t_i$, such as its title and genres. 
Given a set of candidate items $\mathcal C_u = \{c_1, c_2, \ldots, c_M\}$, the goal is to generate a ranked list $\bm {\pi}_u = [\pi_{u,1}, \pi_{u,2}, \ldots, \pi_{u,M}]$, and estimate the confidence $b_{u,p} \in [0,1]$ for each position $p \in \{1, \ldots, M\}$.
\vspace{-0.2mm}
\subsection{Evidence-Based Confidence Estimation}
\label{Sec:conf}
We propose a three-source evidence aggregation framework to estimate position-level confidence. Our key insight is that  the model's decision-making process should reflect a reliable confidence. 
We therefore extract three types of evidence from a single forward pass of the LLM: semantic evidence captures whether the model understands the user's preferences, attention evidence reveals how the model makes decisions, and output evidence quantifies the model's certainty in its final choices. By fusing the evidences, we obtain robust confidence estimates without the computational overhead of multiple sampling approaches.
\vspace{-0.2mm}
\subsubsection{\textbf{Evidence Extraction}}
\vspace{-0.2mm}
\paragraph{\textbf{Semantic Evidence}}
The pairwise ranking theory~\cite{ranknet} shows that a larger score margin between consecutive items corresponds to a lower probability of ranking error. A larger semantic gap between consecutive positions indicates that the model can clearly distinguish between items based on user preferences, suggesting higher confidence in the ranking at this position. Conversely, small gaps suggest ambiguous preference distinctions, reflecting uncertainty.  Therefore, we define semantic evidence as:
\begin{equation}
\label{eq:e_sem}
	e^{sem}_{u,p} = \max \left(\text{sim}(\bm h_{u,p}, \bm c_u ) - \text{sim}(\bm h_{u,p+1}, \bm c_u), 0 \right),
\end{equation}
where $\bm{h}_{u,p}$ is the hidden state at position $p$ from the final transformer layer, $\mathrm{sim}(\cdot,\cdot)$ denotes cosine similarity. The $\max(\cdot, 0)$ ensures evidence is positive only when the item at position $p$ is more relevant than the item at position $p{+}1$, avoiding spurious signals when both items are similarly irrelevant. $\bm{c}_u \in \mathbb {R}^d $ is the user preference context, computed as the mean of historical item embeddings:
\begin{equation}
	\bm c_u = \frac{1}{|\mathcal H_u |} \sum\limits_{i \in \mathcal H_u} \bm i,
\end{equation}
where $\bm i$ represents the item embedding. 
Overall, the semantic evidence for user $u$ as $\bm e^{sem}_u = [e^{sem}_{u,1}, \ldots ,e^{sem}_{u,M}]$.

\paragraph{\textbf{Attention Evidence}}
We examine how the model retrieves information from the input when making its decision at position $p$. If attention is spread nearly uniformly across all input tokens, the model fails to identify relevant context~\cite{voita2019}, suggesting an unreliable decision. We use attention entropy to quantify the degree of concentration or dispersion in the attention distribution~\cite{ghader2017}. To compare sequences of different lengths, we normalize by the maximum entropy:
\begin{equation}
\label{eq:e_att}
	e^{att}_{u,p} = 1 -\frac{H_{u,p}}{\log_2 L}  ,
\end{equation}
where $H_{u,p} = -\sum_{l=1}^{L} a_{u,p}^l \log_2 a_{u,p}^l$ is the attention entropy at position $p$, and $L$ is the number of input tokens.
Low entropy (concentrated attention) indicates the model focuses on specific informative features, reflecting a clear and justified decision basis. High entropy (diffused attention) suggests the model lacks a coherent rationale, indicating high uncertainty. We denote the attention evidence for user $u$ as $\bm e^{att}_u = [e^{att}_{u,1}, \ldots ,e^{att}_{u,M}]$.

\paragraph{\textbf{Output Evidence}}

While semantic and attention evidence reveal the model's internal reasoning, they do not directly capture the model's final certainty about its choices. Prior work shows that higher maximum generation probability indicates lower prediction uncertainty~\cite{hendrycks2017}. However, the probability at a single position is influenced by the candidate set, making cross-position comparisons unreliable. Therefore, we use the difference in maximum probabilities between consecutive positions to compute output evidence: 
\begin{equation}
\label{eq:e_out}
	e^{out}_{u,p} = |\hat p_{u,p} -\hat p_{u,p+1}|,
\end{equation}
where $\hat p_p $ is the maximum probability over candidate tokens at position $p$.
Large probability differences indicate the model is decisively more certain about one position than the other, suggesting high confidence.  Small difference reflect hesitation between consecutive ranking decisions. We denote the output evidence for user $u$ as $\bm e^{out}_u = [e^{out}_{u,1}, \ldots ,e^{out}_{u,M}]$.


\subsubsection{\textbf{Belief Mass.}}
After extracting evidence from three complementary sources, we need to transform these original evidence values into calibrated belief masses that represent our confidence in each ranking position. We elaborate on the subjective logic~\cite{SL}, which provides a framework to reason about uncertainty by connecting evidence to belief mass through the Dirichlet distribution. 
For the evidence $\bm e^{v}_u$ from view $v$\footnote{$v=1$ denotes semantic evidence, $v=2$ denotes attention evidence, and $v=3$ denotes output evidence.},  
subjective logic establishes a connection between the observed evidence and the parameters of a Dirichlet distribution. Specifically, we induce the Dirichlet parameters as:
\begin{equation}
	\alpha^{v}_{u,p} = e^{v}_{u,p}+1,
\end{equation}
where the additive constant 1 serves as an uninformative prior, ensuring $\alpha^v_{u,p} > 0$ even when no evidence is observed . From the Dirichlet parameters, we derive two types of probability masses :
\begin{itemize}[leftmargin=*]
\setlength{\itemsep}{0pt}
\setlength{\parsep}{0pt}
\setlength{\parskip}{0pt}
\item \textbf{belief mass} $b^v_{u,p}$: represents the reliable probability based on evidence source $v$ at position $p$; 
\item \textbf{overall uncertainty mass} $o^v_u$: represents the portion of probability that remains unassigned due to insufficient evidence.
\end{itemize}
These mass values are computed as:
\begin{equation}
	b^v_{u,p} = \frac{e^v_{u,p}}{S^v_u}, \quad o^v_u = \frac{M}{S^v_u},
\end{equation}
where $S^v_u = \sum\limits_{p=1}^M (e^v_{u,p}+1) = \sum\limits_{p=1}^M \alpha^{v}_{u,p}$ is the Dirichlet strength, which indicates the evidence has been collected across all positions. These mass values are all non-negative and their sum is one
:
\begin{equation}
	o^v_u + \sum\limits_{p=1}^M b^v_{u,p} = 1.
\end{equation}
Positions with more evidence receive proportionally higher belief mass, while the overall uncertainty decreases inversely with total evidence. Overall, for the view $v$, we obtain the independent sets of probability masses $\bm \omega^v_u = (\bm b^v_u , o^v_u) = \left ( \{ b^v_{u,p}\}_{p=1}^M, o_u^v \right ) $ .

\subsubsection{\textbf{Reliable Opinion Aggregation.}}
 After obtaining independent masses from three evidence sources, we fuse them into a unified confidence estimate. A key challenge is that different evidence sources have varying reliability, so we propose a reliable opinion aggregation that automatically adapts fusion weights based on uncertainty of each source.
\begin{definition}
\label{D1}
\textbf{(Reliable Opinion Aggregation)} 
Let $\bm \omega^A = (\bm b^A , o^A)$ and $\bm \omega^B = (\bm b^B , o^B)$ be the probability masses of view $A$ and $B$ over the same ranking, respectively. The reliable aggregated opinion $\bm \omega^{A \underline{\Diamond} B} $  is calculated in the following manner: 
\begin{equation}
\begin{aligned}
	\bm \omega^{A \underline{\Diamond} B} &= \bm \omega^A \underline{\Diamond} \bm \omega^B = (\bm b^{A \underline{\Diamond} B}, o^{A \underline{\Diamond} B} ),\\
	b^{A \underline{\Diamond} B}_p &= \frac{r^Ar^B b^A_p b^B_p +r^A b^A_p o^B+ r^B b^B_p o^A}{1-C},\\
	o^{A \underline{\Diamond} B} &= \frac{(r^Ao^A+o^A)(r^Bo^B+o^B)}{1-C},
\end{aligned}
\end{equation}
where $C = \sum\limits_{i \neq j} r^Ar^B b^A_i b^B_j $ measures conflict between the two masses, and $r^* = 1-o^*$ is a reliable coefficient which computed from the uncertainty mass.
\end{definition}

The $\bm \omega^{A \underline{\Diamond} B}$ represents the aggregation of the independent view of $A$ and $B$. Essentially,
the aggregation rule has properties: (i) \textbf{Reliable weighting}: the belief mass is reinforced only when both sources are highly reliable; (ii)\textbf{ Uncertainty accumulation}: the overall uncertainty accumulates not just the original uncertainties but also the distrust from both sources.

\paragraph{\textbf{Constraint Preservation}}
A critical requirement of Dempster-Shafer theory is that belief masses and uncertainty must form a valid probability distribution, i.e., $o^{A \underline{\Diamond} B} + \sum_p b^{A \underline{\Diamond} B}_p = 1$. We verify that our combination rule preserves this fundamental property.
\begin{equation}
\label{eq:16}
\begin{aligned}
&o^{A \underline{\Diamond} B} + \sum_p b^{A \underline{\Diamond} B}_p \\
&= \frac{1}{1-C} \Bigg[(r^Ao^A+1-r^A)(r^Bo^B+1-r^B)  \\
&+\left.\sum_p (r^Ar^B b^A_p b^B_p +r^A b^A_p o^B+ r^B b^B_p o^A)\right], \\
&= \frac{1}{1-C}\Bigg[(r^Ao^A+1-r^A)(r^Bo^B+1-r^B) \\
&+\left. r^Ar^B\sum_p b^A_p b^B_p +r^Ao^B\sum_p b^A_p + r^Bo^A\sum_p b^B_p\right].
\end{aligned}
\end{equation}
And we have 
\begin{equation}
\begin{aligned}
\small
	\sum_p b^A_p = 1-o^A = r^A,  \quad \sum_p b^B_p = 1-o^B = r^B.
\nonumber
\end{aligned}
\end{equation}
Then
\begin{equation}
\begin{aligned}
&o^{A \underline{\Diamond} B} + \sum_p b^{A \underline{\Diamond} B}_p =\\
&\frac{1}{1-C}\Bigg[(r^Ao^A+1-r^A)(r^Bo^B+1-r^B) \\
&+\left. r^Ar^B\sum_p b^A_p b^B_p + r^Ar^Bo^B + r^Ar^Bo^A\right].
\end{aligned}
\end{equation}
By the identity 
\begin{equation}
(\sum_p b^A_p)(\sum_p b^B_p) = \sum_p b^A_p b^B_p + \sum_{i \neq j} b^A_i b^B_j,
\nonumber
\end{equation}
we have:
\begin{equation}
\sum_p b^A_p b^B_p = r^Ar^B - \sum_{i \neq j} b^A_i b^B_j = r^Ar^B - \frac{C}{r^Ar^B}.
\end{equation}
Substituting and expanding the numerator yields $1-C$, thus:
\begin{equation}
o^{A \underline{\Diamond} B} + \sum_p b^{A \underline{\Diamond} B}_p = \frac{1-C}{1-C} = 1.
\end{equation}
Furthermore, as established in subjective logic framework~\cite{SL}, the fusion operator $\underline{\Diamond}$ is both commutative ($\omega^{A \underline{\Diamond} B} = \omega^{B \underline{\Diamond} A}$) and associative ($\omega^{(A \underline{\Diamond} B) \underline{\Diamond} C} = \omega^{A \underline{\Diamond} (B \underline{\Diamond} C)}$), ensuring that the aggregation result is invariant to the order in which the three evidence opinions are combined.

Following Definition \ref{D1} , we can fuse the final joint opinions  $\bm \omega_u$ from three views with the following rule:
\begin{equation}
\label{eq:bf}
	\bm \omega_u = \bm \omega_u^1 \underline{\Diamond} \bm \omega_u^2  \underline{\Diamond} \bm \omega_u^3 = (\bm b_u, o_u).
\end{equation}
According to the above fusion rules, we can get the final joint masses, and thus get the final probability of each position and the overall uncertainty.

\subsection{Position-aware Calibration}
\label{Sec:cali}

Although reliable opinion aggregation provides theoretically belief estimates, the original belief mass $b_{u,p}$ treats all positions equally. Intuitively, correctly placing an item at the top of the list matters far more than doing so near the bottom. 
We therefore introduce a position-aware calibration function.
Such a calibration function must satisfy two key properties: 
\noindent{(i)~\textbf{nonlinearity}},
 to differentially amplify scores across ranges, thereby sharpening the flat distribution and correcting under-confidence; 
 \noindent{(ii)~\textbf{monotonicity}},
 to preserve the relative ordering implied by the original evidential-positional signal.

Specifically, the position-aware calibration is as follows:
\begin{equation}
	\hat b_{u,p} = \sigma \left(\beta \cdot (b_{u,p} \cdot N_{p})+ \gamma \right),
\end{equation}
where $N_p = 1/\log_2(p+1)$ is the NDCG discount factor representing position importance, $\beta$, $\gamma$ are learnable parameters, and $\sigma(\cdot)$ is the sigmoid function.

Classical calibration methods~\cite{label,TS1999} partition predictions into bins and measure within-bin accuracy, which requires sufficient and balanced positive and negative samples per bin. In recommendation ranking, however, each candidate set contains exactly one ground-truth item among many negatives, making bins too sparse for accuracy estimation. Moreover, we argue that confidence in a ranking should not be interpreted as a correctness probability (as in classification), but rather as a \textbf{position reliability score}. A higher confidence at position $p$ should indicate that the ground-truth item is more likely to reside near the top of the list. This is inherently a ranking-level criterion rather than a point-wise probabilistic one. 
Motivated by this observation, we adopt a ranking-adapted calibration objective that supervises ranking confidence to match the NDCG discount target:
\begin{equation}
	\mathcal L_{calib} = \sum\limits_{i^* \in \mathcal C_u} (\hat b_{u,p^*} - N_{p^*})^2,
\end{equation}
where $i^*$ is the ground truth item appearing at position $p^*$ in the LLM ranking. 
We emphasize that this formulation does not claim to produce calibrated probabilities in the classical sense; instead, it produces {calibrated ranking confidence} whose relative magnitudes faithfully reflect position-level reliability.

\vspace{-0.3mm}
\subsection{Confidence-guided Reranking}
\label{Sec:opti}
The calibrated confidence $\hat b = [b_1, \ldots ,b_M]$ provides position-level reliability estimates, but solely re-ranking by confidence would ignore valuable signals from the original LLM ranking scores. Therefore, we propose an optimized scoring function:
\begin{equation}
	s_{u,p} = \hat b_{u,p} \cdot \frac{\exp(l_{u,p})}{\sum_{q \in \mathcal C_u} \exp(l_{u,q})},
\end{equation}
where $l_{u,p}$ is the output logits for item $i_p$ from the final layer of LLM.
We optimize the scoring function using Bayesian Personalized Ranking~\cite{BPR}, which encourages the ground truth item to be ranked higher than all other items:
\begin{equation}
	 \mathcal L_{rank} = -\sum\limits_{\substack{q \in \mathcal C_u\\ q \neq p*}} \log \sigma (s_{u,p^*} - s_{u,q}).
\end{equation}
The final objective function  is given by:
\begin{equation}
	\mathcal L =  \mathcal L_{calib}+ \lambda \mathcal L_{rank},
\end{equation}
where $\lambda$ is a hyper-parameter to balance the tasks.

\vspace{-0.2cm}
\section{Experiment}
\subsection{Experimental Settings}

\paragraph{\textbf{Dataset}}
We conduct experiments on three widely used public datasets for sequential recommendation: MovieLens 1M~\cite{ml1m}, Amazon Grocery~\cite{amazon} and Steam~\cite{steam}. 
The three datasets vary considerably in scale and domain characteristics, enabling a comprehensive evaluation of our method across different recommendation scenarios.
Following previous work~\cite{kweon2025uncertainty,llm4rerank}, we filter out users and items with fewer than 5 interactions. 
The statistics of datasets are summarized in Table \ref{tab:dataset}. 
\begin{table}[h]
    \centering
\setlength{\tabcolsep}{2.5 pt}
\renewcommand{\arraystretch}{1.2}
\caption{Statistics of datasets.\label{tab:dataset}}
\vspace{-0.5mm}
\resizebox{0.95\linewidth}{!}{
    \begin{tabular}{lccc}
        \toprule  
        {\textbf{Dataset}} &
        \textbf{MovieLens 1M} &\textbf{Amazon Grocery} &\textbf{Steam} \\
        \midrule
         {Users} & 6,040 & 21,027 & 38,503 \\
         {Items} & 3,883 & 18,857 & 6,267 \\
         {Ratings} & 1,000,207 & 358,602 & 1,722,038 \\
         {Sparsity} & 95.74\% & 99.91\% & 99.29\% \\
        \bottomrule
    \end{tabular}}
\end{table}

\vspace{-0.3cm}
\paragraph{\textbf{Compared Methods}}
We evaluate our proposed approach across two evaluation scenarios:
\begin{itemize}[leftmargin=*]
\setlength{\itemsep}{0pt}
\setlength{\parskip}{1pt}
    \item \textbf{Recommendation}: We evaluate the recommended performance of our proposed method through  comprehensive baselines. These include traditional sequential recommenders like GMF~\cite{gmf}, SASRec~\cite{steam} and BERT4Rec~\cite{bert4rec}, along with several recent LLM-based approaches such as PepRec~\cite{peprec}, RankGPT~\cite{rankgpt}. Our primary comparison is against LLM4Rerank~\cite{llm4rerank}, which represents the current state-of-the-art for this task.
 
    \item \textbf{Uncertainty Quantification}: We adopt various methods to compute the total predictive uncertainty: Label Prob.~\cite{label}, Semantic Unc.~\cite{semantic}, and Verb. 1S top-1~\cite{verb.1S}. All uncertainty methods are evaluated on the same LLM backbone for fair comparison.
\end{itemize}

\begin{table*}[ht]
\centering
\setlength{\tabcolsep}{4pt}
\renewcommand{\arraystretch}{1.25}
\caption{Overall performance comparison for recommendation.}\label{tab:overall_rec}
\vspace{-1.5mm}
\begin{tabular}{cccccc|cccc|cccc}
\toprule
\multicolumn{2}{c}{\multirow{1}{*}{Dataset}} & \multicolumn{4}{c}{MovieLens 1M} & \multicolumn{4}{c}{Amazon Grocery} & \multicolumn{4}{c}{Steam}  \\ 
\multicolumn{2}{c}{Method} & R@5 & N@5 & R@20 & \multicolumn{1}{c}{N@20} & R@5 & N@5 & R@20 & \multicolumn{1}{c}{N@20} & R@5 & N@5 & R@20 & N@20  \\ \midrule
\multirow{3}{*}{\rotatebox{90}{CF}}
&\multicolumn{1}{l|}{GMF}      & 0.4015 & 0.2978 & 0.5012 & 0.3278 & 0.3692 & 0.2605 & 0.4571 & 0.3169 & 0.4152 & 0.3079 & 0.5490 & 0.3824 \\ 
&\multicolumn{1}{l|}{SASRec}   & 0.4286 & 0.3154 & 0.5412 & 0.3759 & 0.3806 & 0.3911 & 0.4980 & 0.3458 & 0.4537 & 0.3407 & 0.5845 & 0.4097 \\ 
&\multicolumn{1}{l|}{BERT4Rec} & 0.4071 & 0.3042 & 0.5316 & 0.3426 & 0.3762 & 0.2775 & 0.4892 & 0.3152 & 0.4372 & 0.3126 & 0.5741 & 0.3734 \\
\multirow{3}{*}{\rotatebox{90}{LLM}}
&\multicolumn{1}{l|}{PepRec}   & 0.4887 & 0.3863 & 0.5614 & 0.4273 & 0.4603 & 0.3492 & 0.5959 & 0.4086 & 0.5325 & 0.4037 & 0.6931 & 0.4831 \\ 
&\multicolumn{1}{l|}{RankGPT}  & 0.4869 & 0.3860 & 0.5611 & 0.4269 & 0.4609 & 0.3496 & 0.5961 & 0.4088 & 0.5330 & 0.4040 & 0.6934 & 0.4833 \\
&\multicolumn{1}{l|}{LLM4Rerank}&0.5174 & 0.4013 & 0.6389 & 0.5616 & 0.5137 & 0.4896 & \textbf{0.7498} & \textbf{0.5904} & 0.5918 & 0.5571 & 0.7944 & 0.6994 \\
\midrule 
\rowcolor{red!5} 
\multicolumn{2}{l|}{\textbf{\textsc{EviRank}(Mistral)}}  & 0.5586 & 0.4110 & 0.6434 & 0.5771 & 0.5130 & 0.4895 & 0.7204 & 0.5857 & 0.6327 & 0.5783 & 0.8387 & 0.6957 \\ 

\rowcolor{red!5} 
\multicolumn{2}{l|}{\textbf{\textsc{EviRank}(Llama3)}}   & \ul{0.5673} & \ul{0.4188} & \ul{0.6810} & \ul{0.5921} & \ul{0.5479} & \ul{0.5021} & {0.7353} & {0.5894} & \ul{0.6497} & \ul{0.5960} & \ul{0.8615} & \ul{0.7178} \\


\rowcolor{red!5} 
\multicolumn{2}{l|}{\textbf{\textsc{EviRank}(Qwen2.5)}}  & \textbf{0.5702}  & \textbf{0.4239} & \textbf{0.6937} & \textbf{0.6018} & \textbf{0.5602} & \textbf{0.5131} & \ul{0.7487} & \ul{0.5901} & \textbf{0.6583} & \textbf{0.6114} & \textbf{0.8793} & \textbf{0.7294} \\ 

\bottomrule
\end{tabular}
\end{table*}
\begin{table*}
\centering
\setlength{\tabcolsep}{4.3pt}
\renewcommand{\arraystretch}{1.2}
\caption{Overall performance comparison for uncertainty.}\label{tab:overall_un}
\vspace{-1.5mm}
\begin{tabular}{clcccc|cccc|cccc}
\toprule
\multicolumn{2}{c}{\multirow{1}{*}{Dataset}} & \multicolumn{4}{c}{MovieLens 1M} & \multicolumn{4}{c}{Amazon Grocery} & \multicolumn{4}{c}{Steam}  \\ 
\multicolumn{2}{c}{Method} & $\tau$@5 & C@5 & $\tau$@20 & \multicolumn{1}{c}{C@20} &$\tau$@5 & C@5 & $\tau$@20 & \multicolumn{1}{c}{C@20} &$\tau$@5 & C@5 & $\tau$@20 & \multicolumn{1}{c}{C@20}   \\ \midrule
\multirow{4}{*}{\rotatebox{90}{Mistral}}
&\multicolumn{1}{l|}{Label Prob.}    & 0.1010 & 0.5631 & 0.0826 & 0.5441 & 0.0993 & 0.5614 & 0.0781 & 0.5398 & 0.1203 & 0.5733 & 0.1035 & 0.5541   \\
&\multicolumn{1}{l|}{Semantic Unc.}  & 0.1089 & 0.5687 & 0.0898 & 0.5469 & 0.1146 & 0.5712 & 0.0901 & 0.5468 & 0.1390 & 0.5842 & 0.1196 & 0.5622  \\
&\multicolumn{1}{l|}{Verb. 1S top-1} & 0.1123 & 0.5617 & 0.0930 & 0.5428 & 0.1023 & 0.0818 & 0.5677 & 0.5435 & 0.1417 & 0.5864 & 0.1205 & 0.5631  \\
\rowcolor[HTML]{F6FDFE} 
&\multicolumn{1}{l|}{\textbf{\textsc{EviRank}}}       & \textbf{0.1510}  & \textbf{0.5932} & \textbf{0.1373} & \textbf{0.5748} & \textbf{0.1658} & \textbf{0.6021} & \textbf{0.1435} & \textbf{0.5817} & \textbf{0.1805} & \textbf{0.6163} & \textbf{0.1662} & \textbf{0.5931} \\
\midrule \midrule
\multirow{4}{*}{\rotatebox{90}{Llama3}}
&\multicolumn{1}{l|}{Label Prob.}     & 0.1283 & 0.5565 & 0.1143 & 0.5343 & 0.1410 & 0.6001 & 0.1328 & 0.5655 & 0.2312 & 0.6517 & 0.2045 & 0.6190  \\ 
&\multicolumn{1}{l|}{Semantic Unc.}   & 0.1162 & 0.5587 & 0.1072 & 0.5380 & 0.1623 & 0.6137 & 0.1472 & 0.5730 & 0.2767 & 0.6803 & 0.2384 & 0.6331  \\
&\multicolumn{1}{l|}{Verb. 1S top-1}  & 0.1369 & 0.5598 & 0.1260 & 0.5471 & 0.1914 & 0.6290 & 0.1640 & 0.5859 & 0.3041 & 0.6922 & 0.2811 & 0.6591  \\
\rowcolor[HTML]{F6FDFE} 
&\multicolumn{1}{l|}{\textbf{\textsc{EviRank}}}        & \textbf{0.1478} & \textbf{0.5805} & \textbf{0.1369} & \textbf{0.5671} & \textbf{0.1907} & \textbf{0.6279} & \textbf{0.1631} & \textbf{0.5830} & \textbf{0.3035} & \textbf{0.6847} & \textbf{0.2810} & \textbf{0.6573} \\
\midrule  \midrule
\multirow{4}{*}{\rotatebox{90}{Qwen2.5}}
&\multicolumn{1}{l|}{Label Prob.}    & 0.1426 & 0.5793 & 0.1205 & 0.5593 & 0.2289 & 0.6453 & 0.1803 & 0.5979 & 0.2141 & 0.6335 & 0.1957 & 0.6092   \\ 
&\multicolumn{1}{l|}{Semantic Unc.}  & 0.1410 & 0.5835 & 0.1172 & 0.5572 & 0.2320 & 0.6542 & 0.1857 & 0.5991 & 0.2125 & 0.6311 & 0.1949 & 0.6070   \\
&\multicolumn{1}{l|}{Verb. 1S top-1} & 0.1482 & 0.5855 & 0.1257 & 0.5597 & 0.2315 & 0.6539 & 0.1846 & 0.5987 & 0.2131 & 0.6324 & 0.1953 & 0.6078   \\
\rowcolor[HTML]{F6FDFE} 
&\multicolumn{1}{l|}{\textbf{\textsc{EviRank}}}       & \textbf{0.1576}  & \textbf{0.6002} & \textbf{0.1489} & \textbf{0.5873} & \textbf{0.2661} & \textbf{0.6739} & \textbf{0.2301} & \textbf{0.6286} & \textbf{0.3133} & \textbf{0.7095} & \textbf{0.2875} & \textbf{0.6587} \\

\bottomrule
\end{tabular}
\end{table*}

\vspace{-0.15mm}
\paragraph{\textbf{Evaluation Metrics}}
We evaluate both recommendation and uncertainty quantification, and report results at $K \in \{5, 20\}$. 
For recommendation, we follow standard practices in sequential recommender by using Normalized Discounted Cumulative Gain (N@$K$)~\cite{NDCG} and Recall (R@$K$) to evaluate ranking quality. 
For uncertainty quantification, we following~\cite{kweon2025uncertainty} adopt Kendall's $\tau$ ($\tau$@$K$)~\cite{tau} and Concordance Index (C@$K$)~\cite{c}  to assess whether a recommendation with lower uncertainty yields higher N@$K$ than one with higher uncertainty.
\vspace{-0.15mm}
\paragraph{\textbf{Implementation Details}}
Following prior work~\cite{kweon2025uncertainty,llm4rerank}, 
we construct a candidate set $\mathcal{C}_u$ for each user $u$ in the test set. Specifically, we first identify their ground-truth next item $i_{gt}$ from held-out interaction data. We then employ GMF~\cite{gmf} as the candidate generator to retrieve $M-1$ negative items that the user has never interacted with. The final candidate set is thus formed by the union $\mathcal{C}_u = \{i_{gt}\} \cup \{i_{neg_1}, \ldots, i_{neg_{M-1}}\}$. This retrieval-based approach better simulates real-world recommendation pipelines where LLMs serve as re-rankers over candidates from upstream models. To mitigate positional bias in LLMs, the order of items in $\mathcal{C}_u$ is randomly shuffled before being incorporated into the prompt. The candidate size is set to $M = 20$. The maximum user history length is set to 20 for MovieLens-1M and Steam, and 10 for Amazon Grocery.
All experiments are conducted using PyTorch on a single NVIDIA A100-80G GPU. 
\vspace{-0.2cm}
\paragraph{\textbf{Base Large Language Models}} We utilize three  publicly available LLMs as the backbone for our recommendation model, including Llama3~\cite{llama}, Mistral~\cite{Mistral}, and  Qwen2.5~\cite{qwen2.5}. These models are designed for following user instructions and generating structured outputs, making them well-suited for ranking tasks.

\vspace{-0.2cm}
\subsection{Overall Performance}

The experimental results validate our core hypothesis, explicit confidence estimation at each ranking position, can simultaneously improve both recommendation quality and uncertainty quantification. 
Our method demonstrates consistent improvements across diverse datasets (movie, grocery, game), different evaluation settings (@5 and @20), and multiple LLM backbones (Mistral, Llama3, Qwen2.5). 
We present the experimental results for the two scenarios.
\paragraph{\textbf{Recommendation}} 
Table \ref{tab:overall_rec} shows the recommendation performance across three datasets. Specifically, we can observe the following points: 
(i) Our method consistently outperforms all baselines across different LLM backbones, which demonstrates the effectiveness of our framework in enhancing ranking quality. 
(ii) LLM-based methods benefit from richer semantic understanding of item descriptions and user preferences, which generally outperform traditional collaborative filtering approaches.
(iii) Our approach yields more significant performance improvements in top positions (R@5, N@5). This is because position-aware calibration assigns higher importance to top-ranked positions and focuses computational resources on positions that matter most for user experience.
\paragraph{\textbf{Uncertainty}} 
Table \ref{tab:overall_un} compares the uncertainty quantification quality across different methods. We have the following observations: 
(i) Our method achieves the overall best  across the vast majority of LLM backbones and datasets, it indicates that our multi-source evidence aggregation provides more reliable confidence estimates than other approaches.
(ii) Uncertainty performance improves with a more stronger backbone, suggesting that larger models produce more stable internal signals, thereby enabling more accurate uncertainty quantification. 
(iii) The baseline methods show unstable performance, as it only considers uncertainty from a specific component (output or internal). In contrast, our method aggregates complementary evidences to achieve robust uncertainty quantification.

\subsection{Component Ablation}
To validate the effectiveness of each component in our confidence estimation framework, we conduct comprehensive ablation studies. 

\paragraph{\textbf{Recommendation}} 
We use  Qwen2.5 as the backbone and define the following ablation variants:
\begin{itemize}[leftmargin=*]
\setlength{\itemsep}{0pt}
\setlength{\parsep}{0pt}
\setlength{\parskip}{0pt}
\item \textbf{w/o Conf (without Confidence-guided Rerank).} we directly uses the LLM's original ranking without confidence-guided reranking.
\item \textbf{w/o Cali (without Calibration).} We remove confidence calibration and use the original belief quality (Eq.~\ref{eq:bf}) directly to guide the LLM’s reordering.
\end{itemize}
\vspace{-0.2cm}
\begin{table}[h]
\centering
\setlength{\tabcolsep}{3.5 pt}
\renewcommand{\arraystretch}{1.26}
\caption{Ablation study on recommendation.}\label{tab:abrec}
\vspace{-1.5mm}
\begin{tabular}{clcccc}
\toprule
Dataset & Method & R@5 & N@5 & R@20 & \multicolumn{1}{c}{N@20}    \\ \midrule
\multirow{3}{*}{\makecell[c]{Movielens \\ 1M}}
&\multicolumn{1}{c|}{w/o Conf} &  0.5035 & 0.3968 & 0.6320 & 0.5594 \\
&\multicolumn{1}{c|}{w/o Cali} &  0.5271 & 0.4013 & 0.6458 & 0.5713 \\
&\multicolumn{1}{c|}{\cellcolor{red!5}\textsc{EviRank}}   & \cellcolor{red!5}\textbf{0.5702} 
& \cellcolor{red!5}\textbf{0.4239} & \cellcolor{red!5}\textbf{0.6937} & \cellcolor{red!5}\textbf{0.6018}\\
\midrule
\multirow{3}{*}{\makecell[c]{Amazon \\ Grocery}} 
&\multicolumn{1}{c|}{w/o Conf}   & 0.5104 & 0.4520 & 0.7173 & 0.5749 \\
&\multicolumn{1}{c|}{w/o Cali}   & 0.5317 & 0.4626 & 0.7288 & 0.5801 \\
&\multicolumn{1}{c|}{\cellcolor{red!5}\textsc{EviRank}}  & \cellcolor{red!5}\textbf{0.5602}&  \cellcolor{red!5}\textbf{0.5131} & \cellcolor{red!5}\textbf{0.7487}&  \cellcolor{red!5}\textbf{0.5901} \\
\midrule

\multirow{3}{*}{{Steam}}
&\multicolumn{1}{c|}{w/o Conf}   & 0.5836 & 0.5282 & 0.7885 & 0.6812 \\
&\multicolumn{1}{c|}{w/o Cali}   & 0.5891 & 0.5316 & 0.7931 & 0.6843 \\
&\multicolumn{1}{c|}{\cellcolor{red!5}\textsc{EviRank}}   &  \cellcolor{red!5}\textbf{0.6583}&   \cellcolor{red!5}\textbf{0.6114}&   \cellcolor{red!5}\textbf{0.8793}&   \cellcolor{red!5}\textbf{0.7294} \\

\bottomrule
\end{tabular}
\vspace{-0.2cm}
\end{table}

Table~\ref{tab:abrec} presents the ablation results of recommendation.
Both components contribute consistently across all three datasets. Removing confidence-guided rerank yields the largest performance drop, confirming that confidence-weighted scoring effectively identifies unreliable ranking decisions and promotes more trustworthy recommendations to top positions. Introducing original belief mass already recovers substantial performance, indicating that evidential confidence provides meaningful ranking guidance even without calibration. The full \textsc{EviRank} model achieves further improvements, demonstrating that both confidence-guided rerank and position-aware calibration are complementary and indispensable for improving recommendation performence.

\begin{figure*}
\centering
\subfigure[Effect of $\lambda$ on Movielens 1M.]{\label{fig:baby-a}
\includegraphics[width=0.31\textwidth]{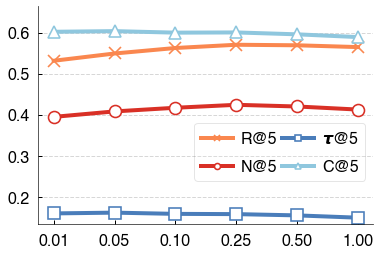}}
\hspace{0.01\textwidth}
\subfigure[Effect of $\lambda$ on Amazon Grocery.]{\label{fig:clothing-a}
\includegraphics[width=0.31\textwidth]{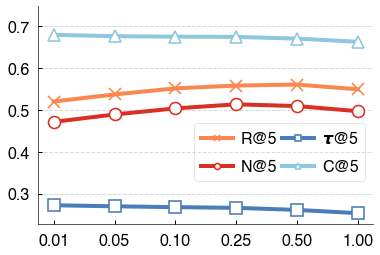}}
\hspace{0.01\textwidth}
\subfigure[Effect of $\lambda$ on Steam.]{\label{fig:kuai-a}
\includegraphics[width=0.31\textwidth]{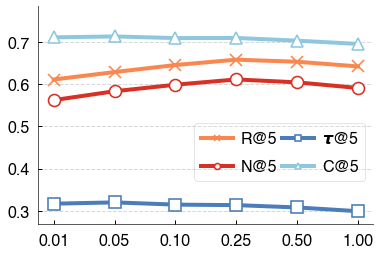}}
\caption{Parameter sensitivity analysis of $\lambda$ (Qwen2.5).} \label{fig:lambda}

\end{figure*}


\paragraph{\textbf{Uncertainty}}

We conduct comprehensive ablation experiments on MovieLens 1M and define the following ablation variants:
%
\begin{itemize}[leftmargin=*]
\setlength{\itemsep}{0pt}
\setlength{\parsep}{0pt}
\setlength{\parskip}{0pt}
\item \textbf{w/o SE (without Semantic Evidence).} We remove the semantic evidence $e^{sem}_{u,p}$ (Eq.~\ref{eq:e_sem}) and aggregate only attention and output evidence.
\item \textbf{w/o AE (without Attention Evidence).} We remove the attention evidence $e^{sem}_{u,p}$ (Eq.~\ref{eq:e_att}) and aggregate only semantic and output evidence.
\item \textbf{w/o OE (without Output Evidence).} We remove the output evidence $e^{sem}_{u,p}$ (Eq.~\ref{eq:e_out}) and aggregate only semantic and attention evidence.
\item \textbf{Only Semantic / Attention / Output.} We use only a single evidence source for belief mass computation, by passing the fusion step entirely.
\item \textbf{w/o ROA (without Reliable Opinion Aggregation).} We apply the standard Dempster's combination rule without our reliability coefficient, treating all evidence sources as equally reliable.
\item \textbf{w/o PaC (without Position-aware Calibration).} We directly use the original fused belief mass (Eq.~\ref{eq:bf}) as the final confidence.
\end{itemize}
%
\begin{table}
\centering
\setlength{\tabcolsep}{6 pt}
\renewcommand{\arraystretch}{1.2}
\caption{Ablation study on uncertainty quantification.}\label{tab:ab}
\vspace{-1.5mm}
\begin{tabular}{clcccc}
\toprule
\multicolumn{2}{c}{Variant} & $\tau$@5 & C@5 & $\tau$@20 & \multicolumn{1}{c}{C@20}    \\ \midrule
\multirow{6}{*}{\rotatebox{90}{Source}}
&\multicolumn{1}{c|}{w/o SE}   & 0.1498 & 0.5941 & 0.1412 & 0.5802 \\
&\multicolumn{1}{c|}{w/o AE}   & 0.1456 & 0.5893 & 0.1387 & 0.5768 \\
&\multicolumn{1}{c|}{w/o OE}   & 0.1552 & 0.5912 & 0.1478 & 0.5861 \\
&\multicolumn{1}{c|}{Only SE}  & 0.1428 & 0.5842 & 0.1198 & 0.5589 \\
&\multicolumn{1}{c|}{Only AE}  & 0.1512 & 0.5874 & 0.1379 & 0.5758 \\
&\multicolumn{1}{c|}{Only OE}  & 0.1287 & 0.5723 & 0.1152 & 0.5498 \\
\midrule
\multicolumn{2}{c|}{w/o ROA}   & 0.1526 & 0.5889 & 0.1481 & 0.5856    \\
\midrule
\multicolumn{2}{c|}{w/o PaC}    & 0.1478 & 0.5923 & 0.1389 & 0.5789   \\
\midrule
\rowcolor[HTML]{F6FDFE} 
\multicolumn{2}{c|}{\textbf{\textsc{EviRank}}}  & \textbf{0.1576}  & \textbf{0.6002} & \textbf{0.1489} & \textbf{0.5873}     \\
\bottomrule
\end{tabular}
\vspace{-0.3cm}
\end{table}
The results presented in Table \ref{tab:ab} clearly reveal the contribution of each component for uncertainty quantification.
First, 
removing any single evidence source leads to performance degradation. 
Attention evidence directly reflect the quality of the model's decision-making process and provides the strongest signal for confidence quantification.  Semantic evidence contributes moderately to confidence quantification, but semantic coherence alone cannot distinguish between deliberate decisions and coincidental matches. Output evidence shows the smallest contribution, it may be due to partial overlap with the ranking logic employed in the final scoring function. The impact ranking is: attention > semantic > output.
Second, 
traditional Dempster’s combination treats all evidence sources as equally reliable, while our reliable opinion aggregation incorporates the reliability coefficient,  it adaptively down-weights sources with high uncertainty, thereby improving robustness against unreliable signals. 
Finally, 
position-aware calibration significantly improves confidence quality.  Positions are inherently unequal in ranking task, uniform confidence estimation fails to reflect this critical asymmetry.  Our position-aware calibration can generate confidence scores that better correlate with ranking quality.

\begin{figure}[t]
\centering
\includegraphics[width=0.49\textwidth]{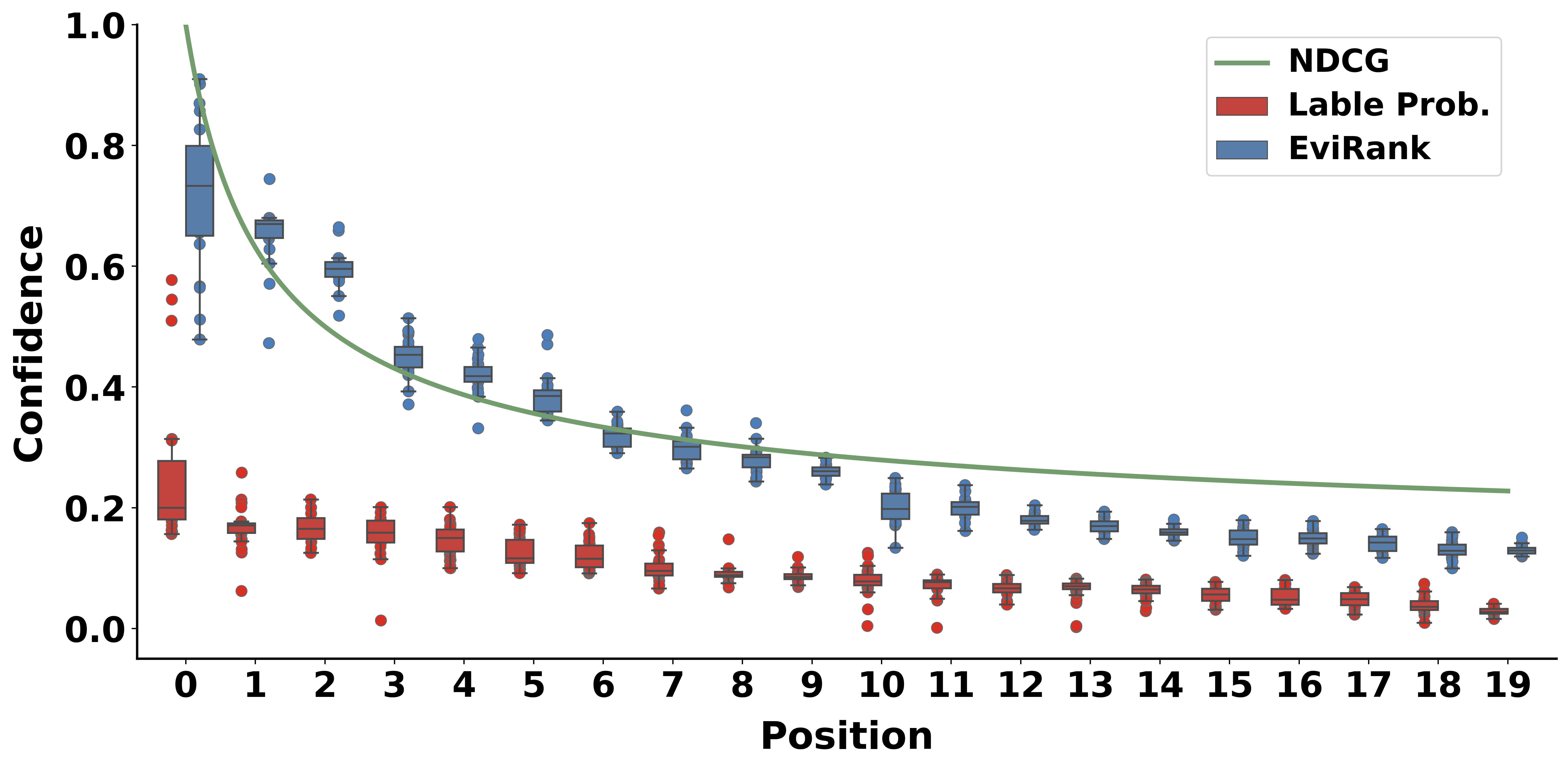} 
\caption{Confidence distributions before and after calibration across rank positions.}
\label{fig:cali}
\end{figure}

\vspace{-0.2mm}
\subsection{Calibration Effectiveness Analysis}
We evaluate two uncertainty quantification methods on the Amazon Grocery. Figure~\ref{fig:cali} shows the confidence score distributions at different positions for the uncalibrated method (Label Prob.) and calibrated method (\textsc{EviRank}), overlaid with the NDCG  curve as reference.
First, the label probability exhibits underconfidence, their median scores cluster near 0.1–0.25 across all positions, which indicates that the model outputs fail to reflect the actual relevance signal at any position. \textsc{EviRank} directly corrects this. The calibrated score distributions are shifted substantially upward at high-relevance positions, closely matching the corresponding NDCG  values. This confirms that position-aware calibration successfully restores the confidence.
Second, and more critically, the label probabilities are nearly flat across all positions, providing no discriminative signal for positional ranking.
The Confidence of \textsc{EviRank} spans a wide range, and the calibrated distribution shows a distinct pattern, particularly among the top 5 most important positions.
In summary, by simultaneously correcting the underconfidence and distributional flatness, EviRank produces well-calibrated confidence scores that can be effectively leveraged to improve performance in downstream tasks such as recommendation.

\vspace{-0.2mm}
\subsection{Hyperparameter Sensitivity}
To further investigate the sensitivity of our model to its key hyperparameters, we analyze the impact of the balancing factor $\lambda$.
The results across three datasets are shown in Figure \ref{fig:lambda}. 
Recommendation performence improves steadily as $\lambda$ increases, while calibration performance degrades gradually beyond the optimal point. 
A small value causes the training objective to be dominated by the calibration loss, which constrains the model to produce conservative  confidence scores at the expense of discriminative reranking. Conversely, a large value over-weights the ranking objective, driving the calibrated scores toward extreme values that improve ranking but lose recommendation reliability. $\lambda = 0.25$ achieves a favorable balance across both tasks on three datasets, where the two loss terms contribute complementarily rather than competitively, allowing the model to jointly refine ranking decisions and confidence estimates. Therefore, we  adopt this value as the default. 

\subsection{Complexity Analysis}
\vspace{-0.6mm}
\begin{table}[h]
    \centering
\setlength{\tabcolsep}{4pt}
\renewcommand{\arraystretch}{1.2}
\caption{Complexity comparison of different methods.\label{tab:ca}}
\vspace{-0.5mm}
{
    \begin{tabular}{lccc}
        \toprule  
        {\textbf{Method}} &
        {Standard LLM} &{LLM4Rerank} &\textsc{EviRank} \\
        \midrule
         \textbf{Time} & ~530.9 ms & ~12,339 ms & ~9,864 ms \\
         
        \bottomrule
    \end{tabular}}
\end{table}

Table~\ref{tab:ca} reports the wall-clock inference time per sample on a single NVIDIA A100-80G GPU using Qwen2.5 on the Amazon Grocery. 
Standard LLM inference without re-ranking serving as the baseline. 
\textsc{EviRank} runs faster than LLM4Rerank, because it extracts all three evidence signals from a single forward pass of the LLM, adding no extra inference steps beyond what reranking requires. 
The aggregation and calibration are lightweight post-processing operations that contribute negligible overhead. Overall, the performance gains of \textsc{EviRank} come at no additional computational cost.

\vspace{-0.2mm}
\section{Conclusion}
Our method \textsc{EviRank} produces position-level confidence estimates by extracting three complementary evidences and aggregating them through reliable opinion fusion. The position-aware calibration further addresses the calibration bias observed in existing methods, enabling effective confidence-guided reranking optimization. Extensive experiments on three datasets demonstrate that \textsc{EviRank} achieves state-of-the-art performance on both recommendation and uncertainty quantification.
For future work, we plan to explore confidence estimation for cold-start scenarios.

\begin{acks}
To Robert, for the bagels and explaining CMYK and color spaces.
\end{acks}

\bibliographystyle{ACM-Reference-Format}
\bibliography{sample-base}


\end{document}